\documentclass[onecolumn,prl,english,superscriptaddress]{revtex4}
\usepackage{amsmath}
\usepackage{natbib}
\usepackage{graphicx}

\begin{document}

\bibliographystyle{apsrev}
\title{Effective charges and virial pressure of concentrated macroion solutions}
\author{N. Boon}
\affiliation{Department of Materials Science and Engineering, Northwestern University, Evanston, Illinois 60208, USA}
\author{G. I. Guerrero-Garc\'ia}
\affiliation{Instituto de F\'isica, Universidad Aut\'onoma de San Luis Potos\'i,
\'Alvaro Obreg\'on 64, 78000 San Luis Potos\'i, San Luis Potos\'i, M\'exico}
\author{R. van Roij}
\affiliation{Institute for Theoretical Physics, Center for Extreme Matter and Emergent Phenomena,  Utrecht University, Leuvenlaan 4, 3584 CE Utrecht, The Netherlands}
\author{M. Olvera de la Cruz}
\affiliation{Department of Materials Science and Engineering, Northwestern University, Evanston, Illinois 60208, USA}

\begin{abstract} The stability of colloidal suspensions is crucial in a wide variety of
processes including the fabrication of photonic materials and scaffolds
for biological assemblies. The ionic strength of the electrolyte that
suspends charged colloids is widely used to control the physical
properties of colloidal suspensions. The extensively used two-body
Derjaguin-Landau-Verwey-Overbeek (DLVO) approach allows for a
quantitative analysis of the effective electrostatic forces between
colloidal particles. DLVO relates the ionic double-layers, which enclose the
particles, to their effective electrostatic repulsion.
Nevertheless, the double layer is distorted at high macroion volume
fractions. Therefore, DLVO cannot describe the many-body effects
that arise in concentrated suspensions. We show that this problem
can be largely resolved by identifying effective point charges for the
macroions using cell theory. This extrapolated point charge (EPC) method
assigns effective point charges in a consistent way, taking into account
the excluded volume of highly charged macroions at any concentration,
and thereby naturally accounting for high volume fractions in both salt-free
and added-salt conditions. We provide an analytical expression for the
effective pair potential and validate the EPC method by comparing molecular
dynamics simulations of macroions and monovalent microions that interact via
Coulombic potentials to simulations of macroions interacting via the
derived EPC effective potential. The simulations reproduce the
macroion-macroion spatial correlation and the virial pressure obtained
with the EPC model. Our findings provide a route to relate the physical
properties such as pressure in systems of screened-Coulomb particles to
experimental measurements.

\end{abstract} 
\date{1 July 2015}
\maketitle

Coulombic interactions between ionized species affect colloidal suspensions at the microscopic level 
and have an indirect, yet crucial, impact on the observable macroscopic characteristics of the system\cite{langmuir1938}. 
The Derjaguin-Landau-Verwey-Overbeek (DLVO) theory \cite{DLVO1, DLVO2}, proposed in the 1940's, 
has been crucial for understanding like-charged colloidal dispersions in a wide variety of 
experimental conditions. In this theory, the effective pair potential between 
two equally charged macroions immersed in an electrolyte
is expressed as the sum of three terms: a hard-core potential that takes into account 
the excluded volume of macroions (preventing their overlap), an 
attractive potential due to short-range (van der Waals) interactions, and 
an electrostatic screened-Coulomb or Yukawa potential resulting from the linearized
Poisson-Boltzmann theory, which is the Debye-H\"uckel approximation.   
Many additions and modifications to the original theory have been proposed, including 
polarization effects, patchiness, or charge regulation, just to mention a few. 
Special care should be taken for non-aqueous solvents, divalent ions, or high salt concentrations, 
since in these regimes ion correlations are usually important
\cite{levine1967,stillinger1960,levine1981,blum1977,valleau1991,oosawa1971,henderson1992,chu1996,netz2000,
panagiotopoulos2002,kjellander2010,wernersson2010,zwanikken2013}.
Generally speaking, modifications to the DLVO theory have been pivotal for systems in which the electrostatics are not well described by the linearized Poisson-Boltzmann theory. 
Though the DLVO theory has been used extensively to model colloidal dispersions\cite{Ninham19991}, this approach is not exact
within the context of the underlying Debye-H\"uckel approximation. In the precise analysis of the force between
two charged spheres in an electrolytic solution, Verwey and Overbeek encountered additional terms that can be considered 
cross terms, resulting from the exclusion of the ionic double-layer surrounding the first sphere by the hard-core of the second sphere\cite{DLVO2}. 
Numerical methods exist to quantify such effects\cite{glendinning1983,carnie1994,warszynski1997}, yet these approaches explicitly deal with two particles in an otherwise empty system.
While in dilute solutions of macroions the resulting correction to DLVO is typically small 
and can usually be safely neglected, in dense macroion systems the deviation from DLVO can become very significant due to the overlap between each macroion electrical double-layer 
with the hard cores of \emph{all} neighbouring particles. As a result, the performance 
of the classical DLVO equation is limited to the description of dilute systems of macroions
\cite{Gruenberg,vanRoij3,warren,Loewen,Hansen}, while many colloidal
processes such as crystallization or glass formation predominantly occur in dense systems
where many-body effects prevail. 
Marcelja et al. recognized the importance of many-body effects at high colloidal densities and low electrolyte
concentration, and they described a method that uses cell theory to project
a charged colloidal dispersion to a system of Coulomb particles\cite{marcelja1976}. 
This enabled use of Monte-Carlo simulations of the Wigner lattice to study the crystallization of latex suspensions. 
That method has recently been rediscovered and extended to charge-regulating particles to describe re-entrant melting
on addition of a charging agent to a colloidal suspension\cite{Kanai2015}. 
Such an effective Coulomb representation is, however, not suitable to describe the structure of charged 
suspensions, particularly in the crystalline phase. 
This is also true of methods comprising the repulsive forces among macroions via hard sphere interactions with effective
hard sphere radii \cite{stigter1954,vanmegen1975,brenner1976,beunen1981}.
Different approaches such as the (renormalized) Jellium model\cite{TrizacLevin} and methods that calculate the osmotic pressure within a 
Wigner-Seitz cell\cite{Alexander,Belloni} have been proposed. However, they do not yield information on the spatial configuration of
the macroions and consequently are limited in describing dense macroion systems.
\\
\indent In this work we introduce a method to calculate
the effective electrostatic pair interaction between macroions in dense systems through the identification 
of their corresponding effective point charges. We verify the corresponding accuracy by comparing the resulting radial distribution 
functions and pressures to the primitive model. To begin, we consider spherical and impenetrable macroions of valence $Z$ 
and radius $a$ immersed in a 1:1 electrolyte with bulk concentration $c_\mathrm{s}$. 
Traditionally for dilute macroion systems, the non-linear Poisson-Boltzmann (PB) theory
establishes that the electrostatic potential is described by 
$\nabla^2 \Phi(\mathbf{r}) = \kappa_\mathrm{res}^2 \sinh \Phi(\mathbf{r})$ outside the macroion, 
where $\Phi(\mathbf{r}) = \Psi(\mathbf{r}) e/k_\mathrm{B}T$, $\Psi(\mathbf{r})$ is the electrostatic potential, 
$e$ is the elementary charge, and $k_\mathrm{B}T$ is the thermal energy of the solution.  
The parameter $\kappa_\mathrm{res} = \sqrt{8 \pi \lambda_\mathrm{B} c_\mathrm{s}}$ is an inverse screening 
length depending on the Bjerrum length $\lambda_\mathrm{B} \equiv e^2/(k_\mathrm{B}T \epsilon)$, where  
$\epsilon$ is the relative dielectric permittivity. For sufficiently small charges,
the PB equation can be linearized by using $\sinh \Phi(\mathbf{r}) \approx
\Phi(\mathbf{r})$, resulting in the Debye-H\"uckel approximation, $\nabla^2 \Phi(\mathbf{r}) =
\kappa_\mathrm{res}^2 \Phi(\mathbf{r})$. The electrostatic potential outside the macroion is found to be 
$\Phi(r) = \lambda_\mathrm{B} Q_{\mathrm{DLVO}}\exp(-\kappa_\mathrm{res} r)/r$,
with $r>a$ the distance to the center of the particle and
$Q_{\mathrm{DLVO}} \equiv Z \exp(\kappa_\mathrm{res} a)/(1+\kappa_\mathrm{res} a)$. The
electric field, and thus the electrostatic force it exerts on a test charge\cite{griffiths1999}, is the same
as that of a point particle with charge $Q_{\mathrm{DLVO}}$. One can therefore
identify $Q_{\mathrm{DLVO}}$ as the effective point charge in the DLVO theory and
estimate the pair potential between two macroions from the screened Coulomb interaction 
of two point charges at a distance $D$,
\begin{equation}
\label{eq:yukawa} 
\frac{U(D)}{k_\mathrm{B}T} =  Q^2 \lambda_\mathrm{B} \frac{\exp(-\kappa D)}{D},  
\end{equation} 
\noindent with $Q = Q_{\mathrm{DLVO}}$ and $\kappa=\kappa_\mathrm{res}$ according to DLVO theory.\\

\begin{figure}[!ht]
\centering\includegraphics[width = 8.5cm]{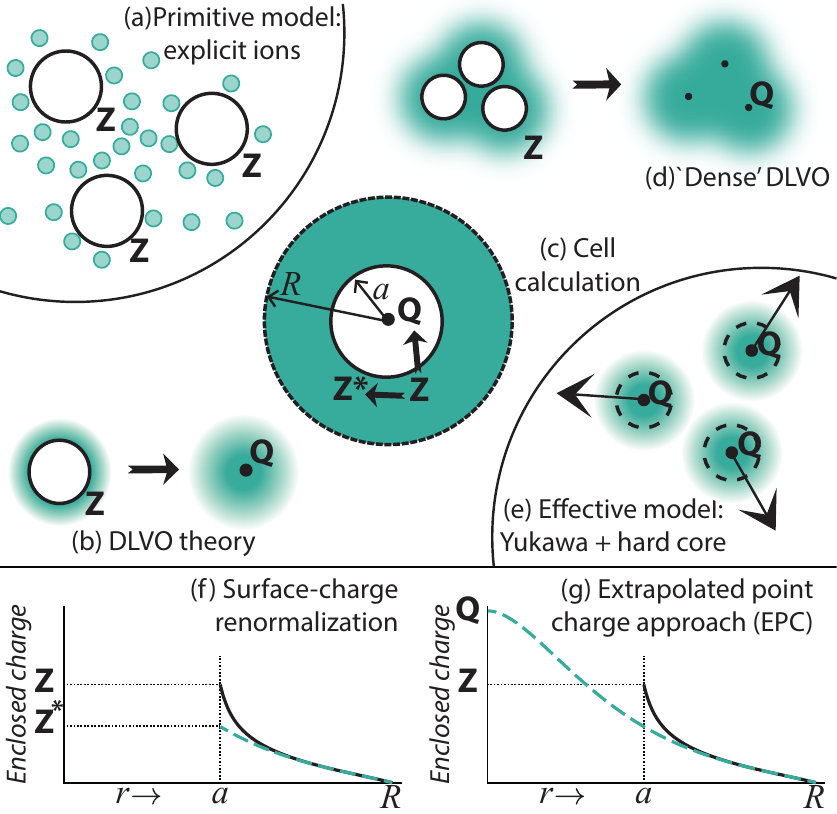} 

\caption{  The various paths (b,c,d) from the primitive model (a) of monovalent
microions and macroions of valency $Z$ to the effective model (e) where
interactions are hard-core Yukawa with effective point charges $Q$. (f) and
(g) show PB-cell calculations for the electric field (or charge within radius
$r$) around a macroion following from nonlinear calculations (full line) and
the Debye-H\"uckel fit (dashed line); (f) illustrates how surface charge renormalization 
yields a charge $Z^*$ that can be inserted into DLVO theory. In the extrapolated point charge approach,
the effective point charge $Q$ is calculated directly from the extrapolation
displayed in (g).\label{fig:routes}} 
\end{figure} 

\indent 
Apart from being restricted to dilute systems, the DLVO equation above cannot directly be applied
to strongly charged macroions, since the Debye-H\"uckel approximation 
no longer holds for these systems, which strictly speaking leads to non-pairwise 
additive interaction potentials \cite{Dobnikar2003,Dobnikar2006}. Alexander and collaborators\cite{Alexander}, 
however, showed that nonlinear ion behavior close to the macroion 
surface can be embodied in an effective linear-screening model by 
calculating a renormalized surface charge $Z^*$ that, far away from
the charged macroion surface, induces 
the same electrostatic potential and electric field as would be obtained within 
the non-linear PB equation \cite{Alexander, Belloni97, Trizac02, Trizac03, Diehl, Denton}, 
see Fig.~1f. Regarding a system of macroions at a concentration $\rho_M$ and macroion 
packing fraction $\eta=4 \pi \rho_M a^3 /3$, 
each of the macroions is imagined to be in the center of a charge-neutral spherical cell with radius
$R=a \eta^{-1/3}$, such that the summed volume of all cells match the system's
volume\cite{Gruenberg}. This is illustrated in Fig.~1c. In this spherical geometry, the
non-linear PB equation and the associated boundary conditions can be written as   
\begin{subequations} 
\begin{align} 
\Phi''(r) + \frac{2 \Phi'(r)}{r} &=
\kappa_\mathrm{res}^2 \sinh \Phi(r);\label{eq:PBequation}\\ 
\Phi'(a) &= -\frac{Z\lambda_\mathrm{B}}{a^2};~~ \Phi'(R) = 0, 
\label{eq:PBequationb} 
\end{align}
\end{subequations} 
\noindent where the prime denotes a derivative with respect to $r$. The boundary conditions
follow from Gauss' law and include the global electroneutrality condition of the whole system.
This set of equations is typically solved numerically as no
general analytical solution is known.  Once the numerical solution is
determined, one proceeds by linearizing Eq.~(\ref{eq:PBequation}) around the
obtained potential at the cell boundary, which can be regarded as the Donnan potential, 
$\Phi_\mathrm{D} \equiv \Phi(R)$. This yields
the Debye-H\"uckel approximation $\Phi_\ell''(r) + 2 \Phi_\ell'(r)/r 
= \kappa^2 \Phi_\ell'(r)$ for the shifted potential $\Phi_\ell(r) \equiv
(\Phi(r) - \Delta\Phi)$, with $\Delta \Phi = \Phi_\mathrm{D} - \tanh \Phi_\mathrm{D}$, and
the screening parameter $\kappa = \kappa_\mathrm{res} \sqrt{\cosh \Phi_\mathrm{D}}$.
Analytical solutions to the linearized PB equation are 
$\Phi_\ell(r) \equiv a_+ e^{ +\kappa r}/r  + a_- e^{-\kappa r}/r$. These form an
accurate approximation to the nonlinear profile in the proximity of the cell's
boundary if one chooses $a_\pm = \exp( \mp \kappa R)\tanh \Phi_\mathrm{D}~ (\kappa R \pm
1)/(2\kappa)$, where the latter follows from the constraints
$\Phi_\ell(R) + \Delta\Phi= \Phi_\mathrm{D}$ and $\Phi'_\ell(R)=0$. The effective
surface charge can now be extracted from the derivative of the analytical
approximation at $r=a$, i.e., $Z^* \equiv -\Phi_\ell'(a) a^2/\lambda_\mathrm{B}$(see Fig.~1(f)),
and one finds $Z^* = a_+/\lambda_\mathrm{B} (\kappa a -1)e^{ +\kappa a} - a_-/\lambda_\mathrm{B} (\kappa a +1)e^{ -\kappa a}$\cite{BZthesis}.
Then, using $\kappa$ and $Z^*$ as parameters, the effective interactions between the
macroions can be estimated by using the DLVO theory again.\\
\indent The accuracy and simplicity of the previous cell-model approach can, however, be improved 
by calculating an effective \emph{point} charge $Q$
\emph{directly} through identification of a point charge at $r=0$ by extrapolating
the analytical approximation(see Fig.~1(g)), yielding the form $Q\equiv\lim_{r\rightarrow 0} -\Phi_\ell'(r)
r^2/\lambda_\mathrm{B} =  (a_+ + a_-)/\lambda_\mathrm{B}$. The latter can also be expressed
as 
\begin{equation}\label{eq:Qrenormalised} Q = \frac{\tanh \Phi_\mathrm{D} }{\kappa
\lambda_\mathrm{B}}\left[\kappa R\cosh \kappa R - \sinh \kappa R\right].  
\end{equation}
The parameters $\kappa$ and $Q$ can then be used to approximate
the effective electrostatic interactions in the original macroion system by 
those of point charges, using Eq.~(\ref{eq:yukawa}) to find the pairwise interaction energy. Although high macroion
volume fractions render DLVO-based approaches inaccurate\cite{Gruenberg,vanRoij3,warren,Loewen,Hansen}, the
effective system of point 
charges has no hard-core volumes that will overlap with ionic double
layers. We therefore expect that Eq.~(\ref{eq:Qrenormalised}) in combination
with Eq.~(\ref{eq:yukawa}) will remain accurate even in dense macroion systems. Note that the
hard-core repulsions for $D<2a$ should be maintained for the non-electrostatic
part of the pair interactions.
Hereafter, we refer to the latter approach as the extrapolated point charge (EPC)
method. The theoretical motivation
for this approach is that screened-Coulomb or Yukawa potentials solve
the screened Poisson equation \emph{without} considering the hard-core
contribution of macroions at finite concentration. Thus the
main advantage of the EPC method is that it defines effective point charges
in a consistent way, taking into account the excluded volume of highly
charged macroions at any concentration.  \\
\indent In the regime where $Z$ is small and the resulting potential profile is 
sufficiently flat throughout the cell,
$|\Phi(R)-\Phi(a)| \ll 1$, the analytical approximations to
Eq.~(\ref{eq:Qrenormalised}) will become exact on the
entire space between the cell boundary and the macroion surface. As a
consequence $a_\pm$ can be calculated from $\Phi_\ell'(a) = -Z\lambda_\mathrm{B}/a^2$
and $\Phi_\ell'(R) = 0$ and a direct analytical relation between $Z$ and $Q$
follows (Fig.~1d). Tantalizingly, inserting this $Q$ into Eq.~(\ref{eq:yukawa})
yields a pair potential similar to the DLVO equation, 
\begingroup \setlength{\medmuskip}{0mu}
\begin{equation}
\label{eq:DDLVO} 
\frac{U(D)}{k_\mathrm{B}T} = \frac{Z^2 \lambda_\mathrm{B} e^{-\kappa
D}\left[2e^{-\kappa (R - a )}(\kappa R \cosh \kappa R~-~ \sinh \kappa
R)\right]^2}{D\left[(1+\kappa a) (\kappa R - 1) ~+~ (1-\kappa a) (\kappa R+1) e^{-2\kappa (R-a)} 
\right]^2}, 
\end{equation} 
\endgroup 

\noindent for $D\geq2a$, and $V(D)
= \infty$ for $D<2a$. The screening parameter $\kappa$ that enters
Eq.~(\ref{eq:DDLVO}) reduces to the reservoir value $\kappa_\mathrm{res}$ for systems with
a sufficient amount of added salt, for which $|\Phi_\mathrm{D}| \ll 1$. Recall that
$R=a\eta^{-1/3}$ and that classical ($\eta$-independent) DLVO theory is re-obtained for
dilute suspensions, which is the limit $R \rightarrow \infty$. 
For completeness we confirm that in the limit of large double-layer size, $\kappa^{-1} \gg D$, Eq.~(\ref{eq:DDLVO}) reduces to the Coulombic form $U(D)/k_\mathrm{B}T = (Z/(1-\eta))^2 \lambda_\mathrm{B} / D$,
with an effective charge $Z/(1-\eta)$ that is larger than the bare charge $Z$ due to the expulsion of
the ionic background from the hard core\cite{marcelja1976,Kanai2015,Russel,Denton2000,Heinen}.
\\
\indent To verify our proposed prescription, molecular dynamic (MD) simulations
of macroion/microion mixtures with particle diameters $d_M=2a=750$~\AA, and $d_+=d_-=3$~\AA, respectively, 
were performed in the constant number, volume, and temperature (NVT) ensemble using the Large-scale Atomic/Molecular Massively Parallel Simulator (LAMMPS) package \cite{Lammps1}. This extreme size-asymmetry between macroions and microions is 
selected to mimic realistic experimental colloidal systems. 
Macroions and monovalent microions, fulfilling the electroneutral condition, were placed inside a cubic 
simulation box of length $L$ under periodic boundary conditions. In the primitive-model (PM) representation
that we applied here, ionic species are represented by repulsive-core spheres with point charges in their centers immersed in a continuous 
solvent \cite{HynninenPM,Ivan2,Ivan3, Ivan5}. The pairwise 
forces among all particles have a short-range repulsive-core potential component, $u^{rc}_{ij}(D)$, and a 
long-ranged Coulombic pair potential contribution, $u^{el}_{ij}(D)/k_\mathrm{B}T= \lambda_\mathrm{B} z_{i} z_{j}/D$, 
where $z_i$ and $z_j$ are the valences associated to particles $i$ and $j$, respectively. 
These interactions are handled properly, using the particle-mesh Ewald technique\cite{darden1993}.
We model the repulsive-core pair potential between a particle of species $i$ and a particle of species $j$, 
separated by a distance $D$, as an impenetrable hard-core $u^{rc}_{ij}(D)=\infty$ for $D \leq \Delta_{ij}$, 
a shifted-truncated Lennard-Jones potential      
$   
u^{rc}_{ij}(D)/k_\mathrm{B}T=4 \left[     
 (\sigma/(D-\Delta_{ij}))^{12} -     
(\sigma/(D-\Delta_{ij}))^{6} \right] + 1
$ for $\Delta_{ij} < D < \Delta_{ij} + 2^{1/6} \sigma$, and by a potential $u^{rc}_{ij}(D)=0$ for   
$D \geq \Delta_{ij} + 2^{1/6} \sigma$, where $\Delta_{ij}$ $=$ $(d_{i}+d_{j})/2-\sigma$ is the   
hard-core diameter. The parameter $\sigma$ regulates the hardness of the repulsive-core   
interactions. To mimic the hard core interaction characteristic of the primitive   
model, $\sigma$ is set equal to $0.1$ nm. We use $\lambda_\mathrm{B}=7.143$~\AA~throughout the text for theoretical
and simulation calculations. 
Additional details of the simulation setup can be found in Refs.~\cite{Ivan2, Ivan3, Ivan5}.\\

\begin{figure}[!ht] 
\centering
\includegraphics[width = 8.5cm]{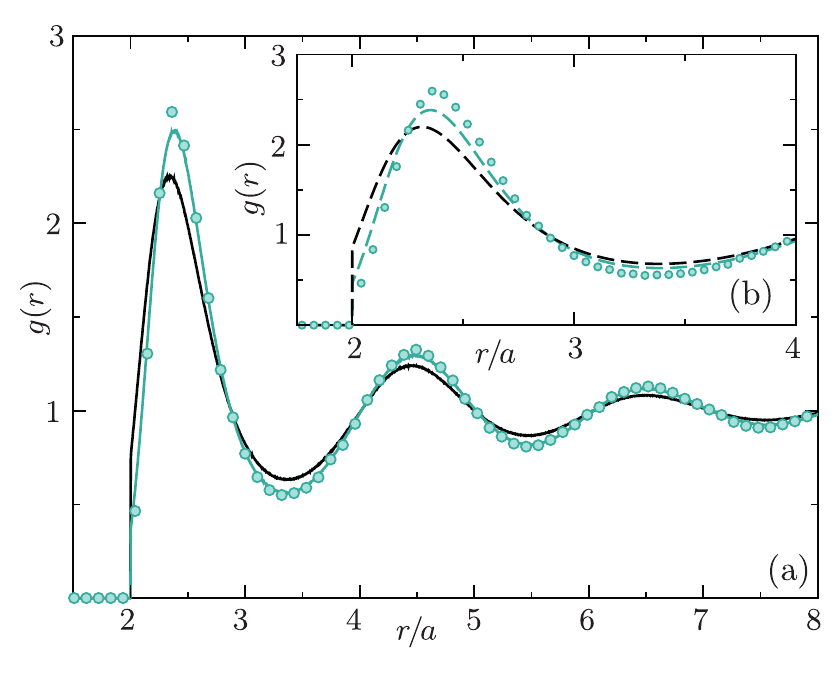}
\caption{Comparison of the pair correlation between macroions resulting
from the full-ion primitive-model MD simulations (circles) with those obtained 
by using repulsive-core effective screened-Coulomb models (lines), relying on the  
the EPC approach (green) and surface charge renormalization approach (black).
The solid lines (a) represent MD simulations results whereas 
the dashed lines in the inset (b) were obtained from the Ornstein-Zernike equation 
within the Rogers-Young closure; both (a) and (b) correspond to the same system in which the valence 
and packing fraction of the macroions in the primitive model are $Z=80$, 
and $\eta=0.3682$, respectively.
\label{fig:alexrenormalization} }
\end{figure}

\indent In Fig.~\ref{fig:alexrenormalization} we compare radial distributions from computationally expensive primitive-model MD simulations (circles)
to much faster and economic effective-model descriptions using MD simulations (main figure, solid lines),  
and integral equations (inset, dashed lines). 
In the primitive-model approach we use a cubic simulation box of length $L=8d_M=6000$~\AA, 
containing 360 macroions of valence $Z=80$, $31680$ small monovalent counterions (-$e$), 
and $2880$ small monovalent co-ions(+$e$). In the effective-model approach, microions 
are included implicitly in the Yukawa interactions between macroions with 
an effective charge $Q$ and inverse screening length $\kappa$. The charges associated to the 
macroion profiles shown in Fig.~\ref{fig:alexrenormalization} are $Q=204$ following the EPC approach and 
$Q=167$ following the surface charge renormalization approach in combination with the DLVO theory.
An excellent agreement between the heavy-duty primitive-model results, in 
which microions are included explicitly, and the computationally inexpensive MD Yukawa 
simulations using the EPC prescription can be observed in Fig.~\ref{fig:alexrenormalization}(a). 
In contrast, surface charge renormalization in combination with the DLVO theory 
deviates significantly from primitive-model simulation results, as expected at this volume fraction. 
The use of integral equations theory allows for an even faster numerical calculation of the radial distribution functions 
within the effective model. 
The Rogers-Young(RY) closure\cite{Rogers}, which is known for its superb accuracy for
hard-core Yukawa systems\cite{Heinen,Castaneda, Gapinski, Zhou}, has good 
qualitative agreement compared with primitive-model results, as 
can be seen in Fig.~\ref{fig:alexrenormalization} (b).
Note that PB techniques are grand canonical and therefore require a reservoir ion density
$c_\mathrm{s}$ or screening parameter $\kappa_\mathrm{res}$, while the number of ions in the
primitive-model system is fixed, as there is no particle exchange with a
reservoir. Therefore, we add an additional step to our PB method to
obtain canonical results: for any choice of $c_\mathrm{s}$ we integrate the resulting ion profiles
in the cell, which yields the total number of ions per macroion. The latter can be compared with the number of
ions per macroion in the simulation box. Subsequently, the right value for
$c_\mathrm{s}$ is determined by a root-finding procedure with respect to their difference. \\

\begin{figure}[!ht] 
\centering\includegraphics[width = 8.5cm]{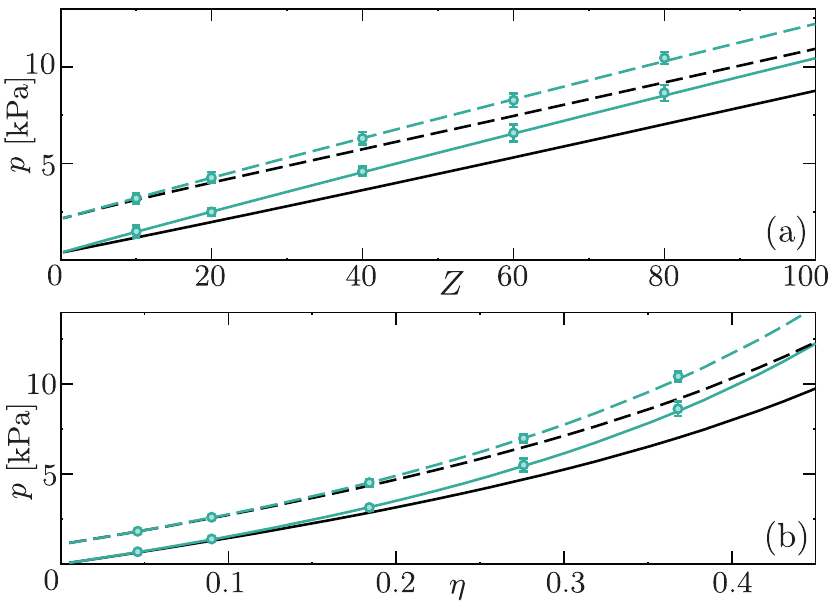}
\caption{The total pressure in the macroion/microion mixture as a function of (a) $Z$
for a fixed macroion volume fraction $\eta=0.3682$, and as a function of (b) $\eta$ and for $Z=80$. 
Both graphs show data for both the salt-free (solid, lower set of lines and
circles) and the added-salt case (dashed, upper set of lines and circles), in which
2880 extra cations and 2880 extra anions were added. The lines result from RY-calculations using the 
effective model parameters following from the EPC approach (green) or surface charge renormalization approach (black). The circles show the results
of primitive-model simulations and error bars are included for those cases where the error
is larger than the circle size. 
\label{fig:pressure} }
\end{figure}

\indent The total microion/macroion pressure resulting from the primitive model, $p_\mathrm{PM}$, as well as the macroion pressure 
in the effective model, $p_\mathrm{EM}$, can be calculated via the virial equation 
$p = \frac{N}{V}k_\mathrm{B}T + \frac{1}{3V}\langle\sum_{i<j}^N D_{ij} \cdot U_{ij}'(D_{ij})\rangle$, 
where $N$ sums all particles, in the primitive model, or only the macroions, in the effective model, 
and $U_{ij}'(D_{ij})$ is the derivative
with respect to the distance $D_{ij}$ of the (un)screened Coulomb pair interaction between particles $i$ and $j$. 
However, to relate  $p_\mathrm{EM}$ to $p_\mathrm{PM}$ it is essential to include a correction
term which can be regarded as the pressure of a homogeneous
background of counterions and co-ions,
\begin{equation}\label{eq:pressurecorrection} 
p_\mathrm{PM} \approx
p_\mathrm{EM} + k_\mathrm{B}T \frac{\kappa^2
}{8\pi\lambda_\mathrm{B}}\left(1+\frac{\kappa_\mathrm{res}^4}{\kappa^4}\right).  
\end{equation} 
Density functional theory\cite{Evans,Gruenberg,Hansen,vanRoij3} may be
applied for a rigorous derivation of this pressure difference, which shows up as a volume term in the free
energy of the effective system\cite{vanRoij3,GrafLoewen,zoetekouw2006,Zoetekouw_PRL}. 
We refer to the supplementary material for this. Note that Eq.~(\ref{eq:pressurecorrection}) 
implies that the macroion osmotic pressure,  which is the pressure with respect to the ion reservoir, 
$\Pi = p_\mathrm{PM} - 2c_\mathrm{s}k_\mathrm{B}T$\cite{Belloni}, does not equal $p_\mathrm{EM}$ in general; only if $\kappa$ 
approaches $\kappa_\mathrm{res}$, the pressure difference in Eq.~(\ref{eq:pressurecorrection}) reduces to $ k_\mathrm{B}T \kappa_\mathrm{res}^2
/(4\pi\lambda_\mathrm{B})=2c_\mathrm{s}k_\mathrm{B}T$.

Fig.~\ref{fig:pressure} shows the primitive-model pressure $p_\mathrm{PM}$ resulting from primitive-model
simulations, as well as the effective-model approximations following from Eq.~(\ref{eq:pressurecorrection})
applying the RY closure. Even though the agreement between the radial distribution functions obtained from
the EPC method using the RY closure and the primitive model simulations is not as accurate as that obtained
from the EPC effective screened Coulomb simulations (see e.g., Fig. \ref{fig:alexrenormalization}), we have
observed that the virial pressure obtained from the RY closure and the effective screened Coulomb
simulations using the EPC approach agreed with the pressure $p_\mathrm{PM}$ obtained from the 
primitive-model simulations within the corresponding numerical uncertainty.
One observes superior accuracy of EPC with respect to 
the surface charge renormalization approach for a wide range in $Z$ (a), and 
in particular for high $\eta$ (b). 
While the dashed lines in Fig.~\ref{fig:pressure} represent added salt cases, 
the full lines correspond to a system without
co-ions, i.e. a salt-free `reservoir':
$c_\mathrm{s}=\kappa_\mathrm{res}^2/(8\pi\lambda_\mathrm{B})=0$ ~\cite{Sainis, Ohshima200218}, for which
Eqs.~(\ref{eq:PBequation}) and ~(\ref{eq:PBequationb}) in principle cannot be
solved. Our cell calculations, however, show that the salt-free limit $\kappa_\mathrm{res}
\rightarrow 0$ is perfectly well defined and can be characterized by the
condition that the number of co-ions
is negligible with respect to the counterions. For small $\kappa_\mathrm{res}$, all relevant physical quantities such as the number of ions 
per macroion, $Q$, and $\kappa$ converge to their values in a salt-free environment. Salt-free cases 
therefore do not require a root-finding procedure
w.r.t $\kappa_\mathrm{res}$, but instead can be considered by choosing $\kappa_\mathrm{res}$ sufficiently
small such that $\kappa_\mathrm{res}/\kappa \ll 1$, i.e. $\Phi_\mathrm{D} \gg 1$. For small macroion
charges, $Z\lambda_\mathrm{B}/a < 1$, we find that the resulting physical $\kappa$ is in
accordance with a homogeneous distribution of neutralizing counterions,
$\kappa^2 = 3 Z \lambda_\mathrm{B} \eta /(a^3~(1-\eta))$~\cite{Gruenberg, Russel,Denton2000,
Ohshima200218}. This, in combination with Eq.~(\ref{eq:DDLVO}), yields a
description at any density of the pair interactions in salt-free systems depending
on $Z$, $\lambda_\mathrm{B}$, $a$ and $\eta$ only. The pressure correction 
in Eq.~(\ref{eq:pressurecorrection}) reduces to
$k_\mathrm{B}T\kappa^2/8\pi\lambda_\mathrm{B}$. The strong dependence of
$\kappa$ on $\eta$, is particularly important
in other geometries, such as the charged two-plate system, from what we study here.
In the latter, the distance between the plates also sets the system's volume
and therefore $\kappa$, yielding a non-exponential form for the
pair-potential~\cite{Briscoe}. However, for macroion suspensions this
is not the case, as $\eta$ is a fixed parameter independent of the
configuration.\\ 
\indent 
In this work we have introduced a method that extends the capabilities of the DLVO theory 
to high valences and volume fractions of colloidal macroions using no additional 
assumptions besides the underlying Poisson-Boltzmann theory. Akin to the original theory, this
approach is mainly applicable to systems with monovalent microions for which ion 
correlations are unimportant, although approximate extensions to systems with correlated 
multivalent counterions might be obtainable.
We also propose a route to relate the pressure in the effective system of macroions to the osmotic
pressure that can be measured experimentally in colloidal systems, for example in sedimentation 
profiles \cite{Torres, Royal}. Our method demonstrates accuracy with respect to
acquiring the measurable properties of charged colloidal suspensions, and can therefore be applied 
to guide and interpret experiments on related systems.

\section{Acknowledgements}
We thank the support of the Center for Bio-Inspired Energy Science (CBES), which is an 
Energy Frontier Research Center funded by the U.S. Department of 
Energy, Office of Science, Basic Energy Sciences under Award DESC0000989. 
{G.I.G.-G. acknowledges the Mexican National Council of Science and Technology (CONACYT) for the financial 
support via the program ``Catedras CONACYT''. } 
The computer cluster where part of the simulations were performed was funded by the Office of the Director of 
Defense Research and Engineering (DDR\&E) and the Air Force Office of Scientific Research (AFOSR) under Award 
no. FA9550-10-1-0167. This work is part of the D-ITP consortium, a program of The Netherlands Organisation for 
Scientific Research (NWO), that is funded by the Dutch Ministry of Education, Culture and Science (OCW).

\bibliographystyle{h-physrev}
\bibliography{bibliography} 
\newpage
\section{Supplementary Material}

Here we will demonstrate how the pressure difference between the effective Yukawa model 
and the primitive model, as stated in
Eq.(5) of the main text, can be derived. We consider a very general charge distribution $q(\mathbf{r})$
without hard core volume, which is in osmotic contact with a reservoir of point ions, 
the latter having a total density of $2c_\mathrm{s}$. Within local mean-field approximation, 
the effective Hamiltonian of such a system can be calculated from the ion profiles
$\rho_\pm(\mathbf{r})$  as [21]

\begin{equation} 
\frac{\Omega}{k_\mathrm{B}T} =
\sum_{\alpha=\pm}\int
\mathrm{d}\mathbf{r}~\rho_{\alpha}(\mathbf{r})\left[\log\frac{\rho_{\alpha}(\mathbf{r})}{c_\mathrm{s}} -
1\right]+\frac{1}{2}\int \mathrm{d}\mathbf{r} \left[\rho_{+}(\mathbf{r}) -
\rho_{-}(\mathbf{r}) + q(\mathbf{r})\right]\Phi(\mathbf{r}). 
\label{eq:PBfunctional}
\end{equation} 
Here, the first term in the integrand can be regarded as the
contribution from the ionic entropy and the second term the electrostatic
energy density of the charge configuration. Note that $\log c_\mathrm{s} = \beta
\mu_{\pm} \Gamma_\pm^{-3}$, with $\mu_\pm$ the ionic chemical potential and
$\Gamma_\pm$ their thermal wavelength. The electrostatic potential is defined
as $\Phi(\mathbf{r}) = \lambda_\mathrm{B} \int \mathrm{d}\mathbf{r}~\left[\rho_{+}(\mathbf{r}) -
\rho_{-}(\mathbf{r}) + q(\mathbf{r})\right]$. Minimization of Eq.~(\ref{eq:PBfunctional})
w.r.t. the ion densities $\rho_\pm(\mathbf{r})$ yields the PB equation, $\nabla
\Phi(\mathbf{r}) = \kappa_\mathrm{res}^2 \sinh \Phi(\mathbf{r}) - 4\pi q(\mathbf{r})$. In
general the latter nonlinear equation cannot be solved analytically; however if
we assume the ion densities to be sufficiently close to a yet undetermined density $\tilde{\rho}$,
then the integrand in Eq.~(\ref{eq:PBfunctional}) can be approximated as a quadratic function of
$(\rho_\pm(\mathbf{r})-\tilde{\rho}_\pm)$. We obtain
\begin{equation} 
\frac{\Omega_\ell}{k_\mathrm{B}T} = -
\frac{V}{2}\left(\tilde{\rho}_+ + \tilde{\rho}_-\right)
+\sum_{\alpha=\pm}\int \mathrm{d}\mathbf{r}~\rho_\alpha(\mathbf{r})\left(\log
\frac{\tilde{\rho}_\alpha}{c_\mathrm{s}} -1\right) +
\frac{1}{2\tilde{\rho}_\alpha }\rho_\alpha(\mathbf{r})^2 
+\int \mathrm{d}\mathbf{r}~ \frac{1}{2} \left[\rho_+(\mathbf{r}) - \rho_-(\mathbf{r}) +
q(\mathbf{r})\right]\Phi(\mathbf{r}), 
\label{eq:PBfunctional_linearized} 
\end{equation} 
with $V$ the system's volume. For a specific macroion/microion mixture,
we determine $\tilde{\rho}_\pm$ and $q(\mathbf{r})$ following the cell 
approach that was described in the main text. The effective point charge $Q$ yields
a charge distribution $q(\mathbf{r}) = \sum_{i\leq M} Q \delta^3 
(\mathbf{r}_i - \mathbf{r})$ for $M$ macroions, and we choose $\tilde{\rho}$ to be the ion densities 
at the cell surface, $\tilde{\rho_\pm} \equiv \rho(R) = c_\mathrm{s} \exp (\mp \Phi_\mathrm{D})$.
As a result Eq.~(\ref{eq:PBfunctional_linearized}) may be transformed into 
\begin{align} 
\frac{\Omega_\ell}{k_\mathrm{B}T} &= - V\frac{c_\mathrm{s}}{\cosh
\Phi_\mathrm{D}} (1 + \cosh^2 \Phi_\mathrm{D}) 
+\sum_{\alpha=\pm}\int \mathrm{d}\mathbf{r}
\frac{\exp (\alpha \Phi_\mathrm{D})}{2c_\mathrm{s}}\left(\rho_\alpha(\mathbf{r}) -
\tilde{c_\mathrm{s}}\right)^2\nonumber\\ 
&+\int \mathrm{d}\mathbf{r}~ \frac{1}{2}
\left[\rho_+(\mathbf{r}) - \rho_-(\mathbf{r}) +
q(\mathbf{r})\right](\Phi_\ell(\mathbf{r})-\Delta\Phi) 
+\int \mathrm{d}\mathbf{r}~q(\mathbf{r}) \Delta\Phi , 
\end{align} 
where $\Delta \Phi = \Phi_\mathrm{D} - \tanh{\Phi_\mathrm{D}}$, $\tilde{c_\mathrm{s}} = c_\mathrm{s} /
\cosh{\Phi_\mathrm{D}}$, and $\Phi_\ell(\mathbf{r}) =
\Phi(\mathbf{r}) - \Delta \Phi$ is the shifted electrostatic potential. From the functional derivatives $\frac{\delta 
\Omega_\ell}{\delta \rho_\pm}=0$ we find
$\rho_\pm (\mathbf{r})= \tilde{c_\mathrm{s}} \mp c_\mathrm{s} \exp(\mp \Phi_\mathrm{D}) \Phi_\ell(\mathbf{r})$
and now the Debye-H\"uckel approximation for the shifted potential $ \Phi_\ell(\mathbf{r})$ is
recovered, $\nabla \Phi_\ell(\mathbf{r}) = \kappa^2 \Phi_\ell(\mathbf{r}) - 4\pi\lambda_\mathrm{B}
q(\mathbf{r})$, with the effective screening parameter $\kappa^2 =\kappa_\mathrm{res}^2\cosh
\Phi_\mathrm{D}$. Our choice is self-consistent since the latter equation reduces to a cell
model in which $\Phi_\ell''(r) + \Phi_\ell'(r)/2r = \kappa^2 \Phi_\ell'(r)$ , with 
$\Phi_\ell'(R)=0$ and $\lim_{r\rightarrow 0} \Phi_\ell'(r) r^2 = -Q \lambda_\mathrm{B}$, describes the
cell potential profiles. This 
reproduces the analytical approximation to the ion profiles within the cell,
and therefore validates our choice of $\tilde{\rho_\pm}$ a posteriori. For the general distribution
of effective point charges, one finds the electrostatic potential $\Phi_\ell (\mathbf{r}) = \sum_{i\leq M} 
Q \lambda_\mathrm{B} \exp(-\kappa |\mathbf{r} - \mathbf{r}_i|)/|\mathbf{r} - \mathbf{r}_i|$, and therefore
\begin{equation}\label{eq:interactionenergy}
\frac{U_\mathrm{EL}}{k_\mathrm{B}T} =\sum_{i<j}
\frac{Q^2\lambda_\mathrm{B}}{|\mathbf{r}_i - \mathbf{r}_j|} \exp (-\kappa |\mathbf{r}_i -
\mathbf{r}_j|) +M(Q \Delta \Phi) -V \frac{\kappa^2
}{8\pi\lambda_\mathrm{B}}\left(\frac{\kappa_\mathrm{res}^4}{\kappa^4} +1\right), 
\end{equation} 
gives the electrostatic energy of the effective macroion system. 
Note that divergent self-energy terms are omitted.  
Now, the electrostatic pressure can be obtained by the usual derivative w.r.t. $V$, 
choosing to keep $\tilde{\rho}_\pm$ constant since those remain
appropriate fixed points in the quadratic approximation to Eq.~(\ref{eq:PBfunctional}) 
for infinitesimal changes of $V$. We obtain

\begin{equation}
p = \left(\frac{\mathrm{d}U_\mathrm{EL}}{\mathrm{d}V}\right)_{M,T,\tilde{\rho}_\pm} = p_\mathrm{Y} + k_\mathrm{B}T\frac{\kappa^2
}{8\pi\lambda_\mathrm{B}}\left(\frac{\kappa_\mathrm{res}^4}{\kappa^4} +1\right).
\end{equation}
Here, $p_\mathrm{Y}$ is the usual electrostatic pressure of the set of point 
charges that interact via an Yukawa interaction, whilst the second term
reveals an additional contribution to the electrostatic pressure.
Therefore, this is the additional pressure that must be added to the total pressure of the effective
(hard-core) Yukawa system to relate to the primitive-model pressure 
which includes both macroions and microions explicitly.\\

\end{document}